\documentclass{ws-procs9x6}

\begin{document}

\title{Hadron spectroscopy at HERA}

\author{M. Barbi}

\address{DESY Laboratory\\
	22607, Hamburg, Germany \\
	and\\
	McGill University\\
        Physics Department, Montreal, Canada\\
E-mail: barbi@mail.desy.de} 

\address{on behalf of the ZEUS and H1 Collaborations}

\maketitle

\abstracts{
Inclusive photoproduction cross-sections of the 
neutral mesons $\eta$, $\rho^0$, $f_0(980)$ and $f_2(1270)$ have been measured 
by H1 and compared to the photoproduction
of $\pi^+$ in $ep$ collisions at HERA. Also, inclusive $K_s^0K_s^0$ 
production and evidence for a narrow baryonic state decaying to
$K_s^0p$ have been observed by ZEUS at HERA.
}

\section{Inclusive photoproduction of $\eta$, $\rho^0$, $f_0(980)$ and
$f_2(1270)$ resonances}

Production of long-lived hadrons at central values of
rapidity in hadron collisions is expected to be independent of
the type of colliding hadrons,
being dominated by the properties of the QCD vacuum. 
At HERA, photoproduction events provide an
opportunity to study particle production in light hadron collisions
at about the same energy as in the heavy ion collisions at RHIC.
In this contribution, the results from the first measurements of 
inclusive photoproduction of 
the resonances $\eta$, $\rho^0$, $f_0(980)$ and
$f_2(1270)$ and a comparison with 
production of particle of other species at $\gamma p$ 
centre-of-mass energy of $\sim$210 GeV using the H1 experiment
are shown~\cite{h1res}. 

A detailed description of the H1 detector can be found elsewhere~\cite{h1det}.
The data used for this analysis correspond to an integrated luminosity of
38 pb$^{-1}$ . 

A small angle positron tagger was used to select photoproduction 
events with photon virtuality $Q^2<10^{-2}$ GeV$^2$. Monte Carlo simulations
were used to estimate the selection efficiency and detector acceptance for 
cross-section measurement. Details on the event selection and the method used
for cross-section calculations can be found in~\cite{h1res}.

The $\eta$ meson candidates were reconstructed
through their $\eta\rightarrow \gamma\gamma$ decay mode using 
the liquid argon calorimeter. 
The $\rho^0$, $f_0(980)$ and $f_2(1270)$ were reconstructed through their
$\pi^\pm\pi^\mp$ decay mode using the central jet chamber. 
The measured photons and charged tracks were required to be 
in the polar angle range $0.5<\theta<2.6$, limiting the study to a region of 
rapidity $|y|<1$ in the laboratory frame.

\begin{figure}
\includegraphics[height=.34\textheight]{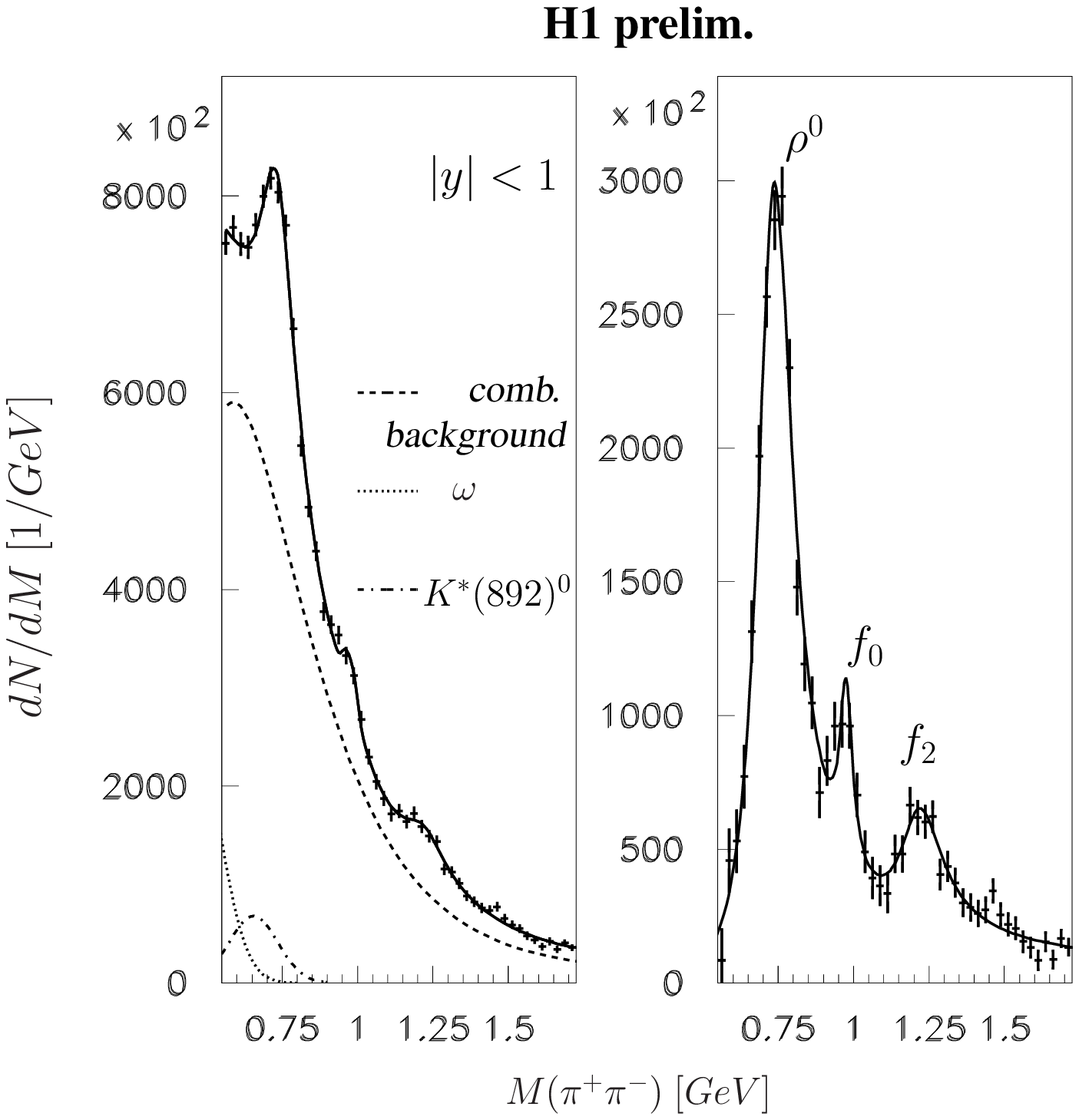}~\includegraphics[height=.3\textheight]{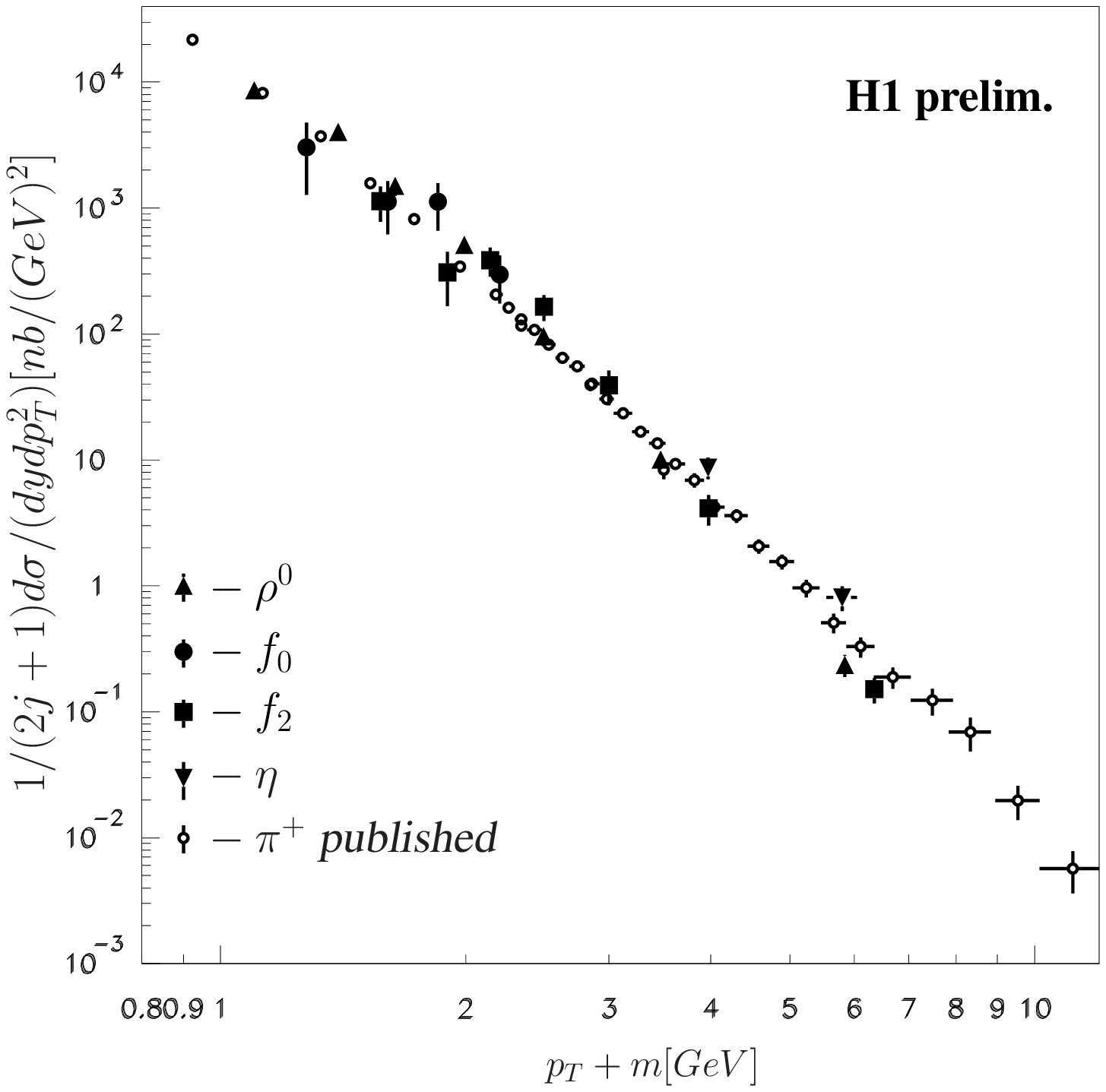}
 \caption{The invariant-masses for the $\rho^0$, $f_0(980)$ and $f_2(1270)$ 
meson candidates are shown before background subtraction (left) and after
background subtraction (center). The differential
cross-sections (right) are shown as a function of $p_T+m$.}
\label{fig:h1invmass}
\end{figure}

Figure~\ref{fig:h1invmass} shows the invariant-mass $M(\pi^+\pi^-)$
distributions for 
the $\rho^0$, $f_0(980)$ and $f_2(1270)$ candidates.
The double differential cross sections for $\eta$, $\rho^0$, $f_0(980)$ and 
$f_2(1270)$ are shown 
as a function of $m+p_T$, where $m$ is the meson nominal mass. Also shown is 
the cross-section for pions~\cite{h1res}
at the same 
$\gamma p$ centre-of-mass energy. 
The cross sections follow a similar power-law function and appear to depend on 
the masses of the hadrons and transverse momentum but not on their 
internal structure. This universal feature for long-lived hadrons is 
supported by~\cite{lonlive}.
These measurements are also important 
to understand hadron production at RHIC, where similar processes
can be wrongly interpreted as associated to formation of a quark-gluon plasma.

\section{Evidence for a narrow baryonic state decaying to $K_s^0p(\bar p)$ in
deep inelastic scattering at HERA}

The existence of a narrow baryon resonance with a mass close
to 1530 MeV and positive strangeness has been reported by several
experiments~\cite{prl91012003}. 
This state has been interpreted as a bound state of 
5 quarks and identified as a candidate for the $\Theta^+$ state ($uudd\bar{s}$)
predicted in the chiral soliton model~\cite{pentamodel}. In this contribution, results 
of a resonance search in the $K^0_Sp(\bar p)$ invariant-mass 
spectrum measured using the ZEUS detector at HERA are presented. 
Details on this analysis can be found in~\cite{pentares}. 

A detailed description of the ZEUS detector can be found 
elsewhere~\cite{zeus:1993:bluebook}.
An integrated luminosity of 121 pb$^{-1}$ was used to select deep inelastic 
scattering events with photon virtuality $Q^2>1$ GeV$^2$ at an $ep$ energy of 
300-318 GeV.

The inclusive DIS selection was defined by requiring an
electron found in the Uranium Calorimeter, and further requirements were applied
to ensure a well defined data sample~\cite{pentares}.

The Central Tracking Detector (CTD) was used to select charged tracks.
The $K_s^0$ candidates were reconstructed through their
$K_s^0\rightarrow\pi^+\pi^-$ decay mode.  
The (anti-)proton candidate was selected using the energy-loss $dE/dx$ measured
in the CTD.
A detailed description of the $K_s^0$ and \mbox{(anti-)proton} candidate 
selection can be found in~\cite{pentares}.

\begin{figure}
\begin{minipage}[t]{48mm}
\centerline{\epsfxsize=2.2in\epsfbox{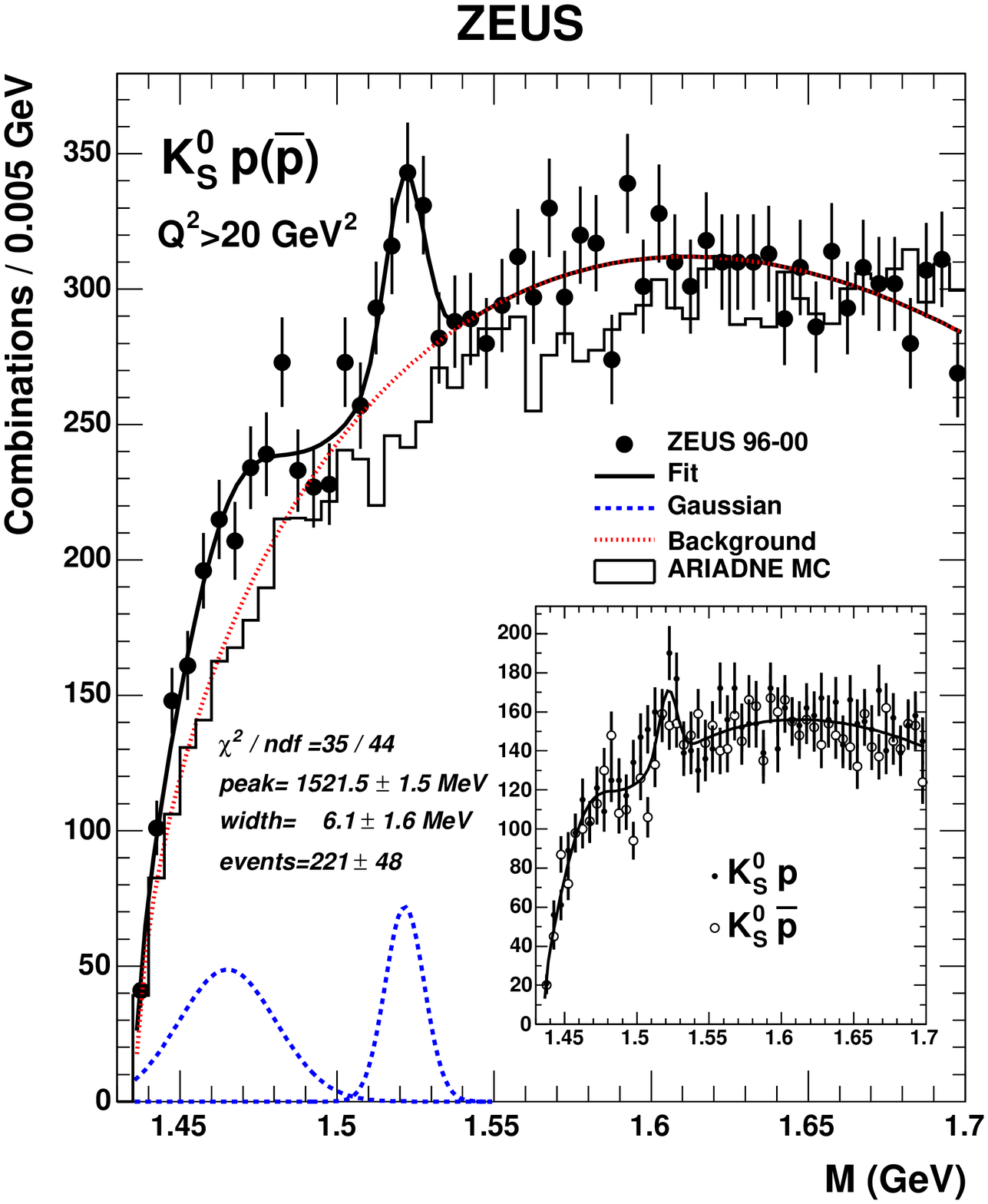}}
\caption{Invariant mass for the $K_s^0p(\bar p)$ resonance. The solid line is 
the result of a fit using two Gaussians (dashed lines) plus a three-parameter 
background function (dotted line). The histogram depicts the predictions from 
the ARIADNE Monte Carlo simulation which contains only well established 
resonances. The inset shows the independent measurements for $K_s^0p$ (black 
dots) and $K_s^0\bar p$ (open circles) candidates.} 
\label{fig:Ksmass} 
\end{minipage} 
\hspace{13mm} 
\begin{minipage}[t]{48mm} 
\centerline{\epsfxsize=2.7in\epsfbox{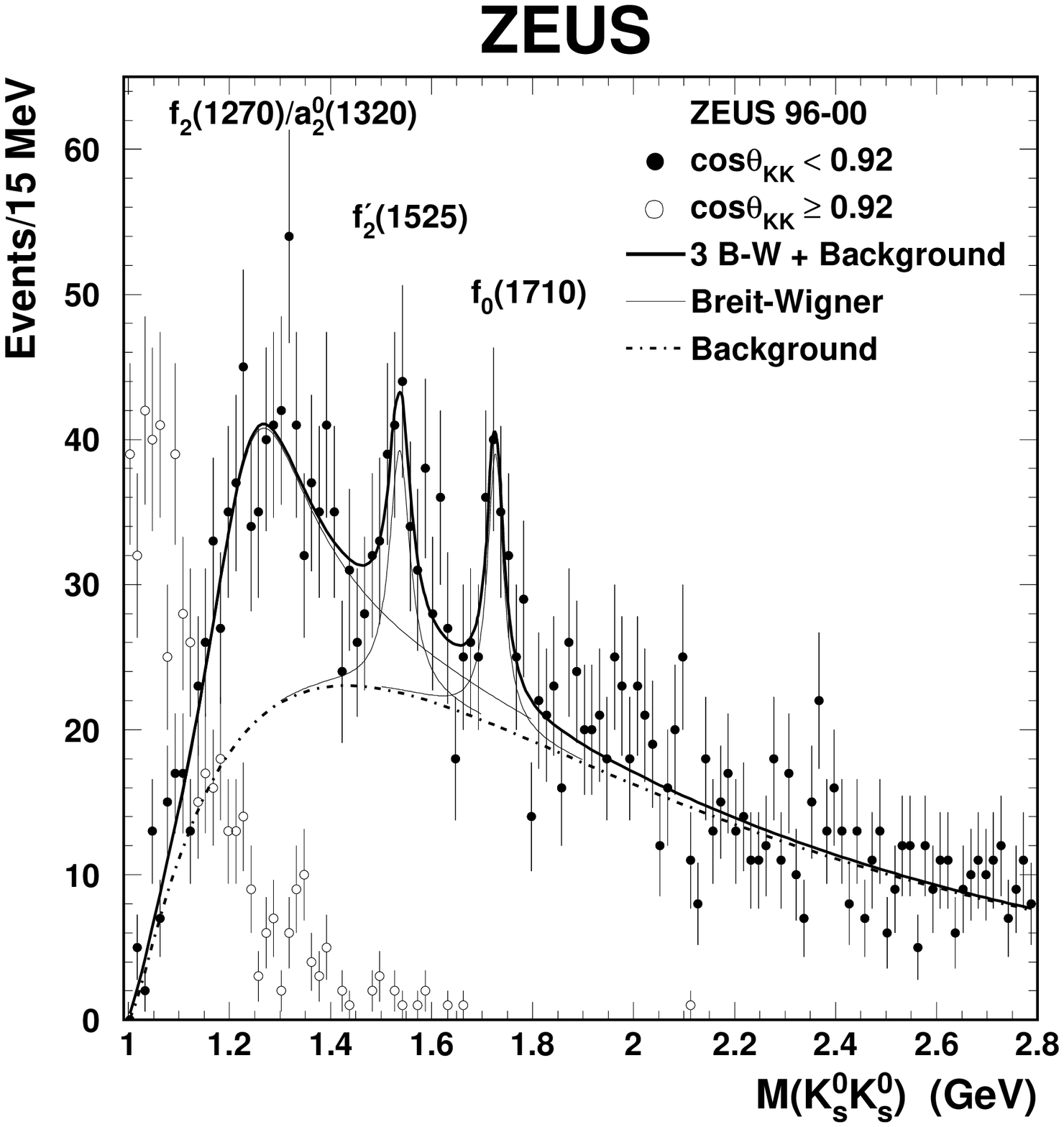}}
\caption{The $K_s^0K_s^0$ invariant-mass spectrum for $K_s^0$ pair candidates 
with $cos\theta_{K_s^0K_s^0}<0.92$ (filled circles).  The thick solid line is 
the result of a fit using three Breit-Wigners (thin solid lines) and a 
background function (dotted-dashed line). The $K_s^0$ pair candidates that 
fail the $cos\theta_{K_s^0K_s^0}<0.92$ cut are also shown (open circles).}
\label{fig:KKmass} 
\end{minipage} 
\end{figure}

The $K_s^0p(\bar p)$ 
invariant-mass spectrum, M,  for\footnote{Distributions in 
different regions of $Q^2$ can be found in~\cite{pentares}} $Q^2>20$ GeV$^2$
is shown in Fig.~\ref{fig:Ksmass}.  
The distribution was fitted using two Gaussians and a
three-parameter background function. 
A peak is seen at 1521.5$\pm$2.9(stat.) MeV with a measured width of
6.1$\pm$1.6(stat.) MeV and significance corresponding to  4$.6\sigma$ 
(3.9$\sigma$ if only one Gaussian is used),
consistent with the predicted $\Theta^+$ pentaquark
with a mass close to 1530 MeV and a width of less than 15 MeV. 
Also shown are the independent measurements for
$K_s^0p$ and $K_s^0\bar p$ candidates. The latter presents the first 
evidence for the production of a
$\bar{\Theta}^+$ ($\bar u\bar u\bar d\bar d s$) state
in a kinematical region dominated by fragmentation processes.

\section{$K_s^0K_s^0$ final state in deep inelastic scattering at HERA}

The $K_s^0K_s^0$ system is expected to couple to scalar
and tensor glueballs~\cite{rmp:v71:1411}. 
Lattice QCD calculations~\cite{glueball1}
predict the existence of a scalar 
glueball with a mass of $1730\pm 100$ MeV which
can mix with $q\overline{q}$ states with $I=0$ from 
the scalar meson nonet, leading to three $J^{PC}=0^{++}$ states whereas
only two can fit into the nonet.
In this contribution, the first observation of resonances in the 
$K_s^0K_s^0$ final state in inclusive deep inelastic $ep$ scattering 
is reported~\cite{hep-0308006}.

Deep inelastic scattering events with $Q^2>4$ GeV$^2$
and with at least one pair of $K_s^0$ candidates
were selected from the same ZEUS data sample 
as described in the previous section. 
The $K_s^0$ candidates were identified through their 
$K_s^0\rightarrow\pi^+\pi^-$ decay mode using the central tracking detector.
A detailed description of the event selection and $K_s^0$ pair 
candidate reconstruction can be found in~\cite{hep-0308006}.

Figure \ref{fig:KKmass} shows the measured $K_s^0K_s^0$ invariant-mass 
spectrum. A strong enhancement near the
$K_s^0K_s^0$ threshold due to the $f_0(980)$/$a_0(980)$
state~\cite{prd:v32:189} was removed by imposing the cut 
\mbox{$cos\theta_{K_s^0K_s^0}<0.92$}, where $\theta_{K_s^0K_s^0}$ 
is the opening angle between the two $K_s^0$ candidates in the 
laboratory frame. Below $1500$ MeV, a region strongly affected by the 
$\cos{\theta_{K_s^0K_s^0}}$ cut, a peak is seen around 
$1300$ MeV where a contribution from $f_2(1270)/a_2^0(1320)$ is expected.
This mass region was fitted with a single Breit-Wigner.

Above $1500$ MeV, the lower-mass state has a fitted mass of 
$1537^{+9}_{-8}$ MeV and a width of $50^{+34}_{-22}$ MeV, in good 
agreement with the well established $f_2^\prime(1525)$ state.
The higher-mass state has a fitted mass of $1726 \pm 7$ MeV and
a width of $38^{+20}_{-14}$ MeV, consistent with the glueball candidate
$f_0(1710)$.

It was found that 93\% of the $K_s^0$-pair candidates selected within the 
detector and trigger acceptance are 
in a region
where sizeable initial state gluon radiation may be expected. 


\bibliographystyle{aipproc}

\bibliography{sample}

\IfFileExists{\jobname.bbl}{}
 {\typeout{}
  \typeout{******************************************}
  \typeout{** Please run "bibtex \jobname" to optain}
  \typeout{** the bibliography and then re-run LaTeX}
  \typeout{** twice to fix the references!}
  \typeout{******************************************}
  \typeout{}
 }

\end{document}